\documentclass[prd,twocolumn,showpacs,superscriptaddress,nofootinbib,preprintnumbers,aps]{revtex4}
\usepackage{color}
\usepackage{latexsym, amssymb, amsmath,color,graphicx}
\usepackage[unicode=true, 
 bookmarks=true,bookmarksnumbered=false,bookmarksopen=false,
 breaklinks=false,pdfborder={0 0 1},backref=false,colorlinks=true]
 {hyperref}

 \makeatletter

\@ifundefined{textcolor}{}
{%
 \definecolor{BLACK}{gray}{0}
 \definecolor{WHITE}{gray}{1}
 \definecolor{RED}{rgb}{1,0,0}
 \definecolor{GREEN}{rgb}{0,1,0}
 \definecolor{BLUE}{rgb}{0,0,1}
 \definecolor{CYAN}{cmyk}{1,0,0,0}
 \definecolor{MAGENTA}{cmyk}{0,1,0,0}
 \definecolor{YELLOW}{cmyk}{0,0,1,0}
 }

\usepackage{hyperref}
\hypersetup{
    colorlinks,%
    citecolor=blue,%
    filecolor=blue,%
    linkcolor=blue,%
    urlcolor=blue
}

\begin{document}

\title{Collinear and Soft Divergences in Perturbative Quantum Gravity}

\preprint{YITP-SB-11-31}

\author{Ratindranath Akhoury}
\email{akhoury@umich.edu}
\affiliation{Michigan Center for Theoretical Physics, Randall Laboratory of Physics, University of Michigan, Ann Arbor, MI 48109-1120, USA}

\author{Ryo Saotome}
\email{rsaotome@umich.edu}
\affiliation{Michigan Center for Theoretical Physics, Randall Laboratory of Physics, University of Michigan, Ann Arbor, MI 48109-1120, USA}

\author{George Sterman}
\email{george.sterman@stonybrook.edu}
\affiliation{C.N. Yang Institute for Theoretical Physics, Stony Brook University, Stony Brook, New York, 11794-3840, USA}

\date{\today}

\begin{abstract}

Collinear and soft divergences in perturbative quantum gravity are investigated to arbitrary orders in amplitudes for wide-angle scattering, using methods developed for gauge theories.    We show that collinear singularities cancel when all such divergent diagrams are summed over, by using the gravitational Ward identity that decouples unphysical polarizations from the S-matrix. This analysis generalizes a result previously demonstrated in eikonal approximation.   We also confirm that the only virtual graviton corrections that give soft logarithmic divergences are of the ladder and crossed ladder type.

\end{abstract}


\maketitle

\section{Introduction}

Infrared divergences in perturbative quantum gravity were investigated long ago in Ref.\ \cite{Weinberg}, where the exponentiation of the singular contributions in ladder and the crossed ladder diagrams was verified by analogy to quantum electrodynamics.  In the scattering of massless particles, or at very high energies, graviton ladder diagrams, like those in QED, also develop collinear singularities, or ``mass divergences".   In contrast to QED, however, collinear singularities turned out to cancel after the summation of all ladders, when treated in eikonal approximation.  The cancellation of the remaining, non-collinear soft-graviton divergences between virtual and real ladder emission processes was also pointed out in \cite{Weinberg}, and subsequently confirmed in full quantum gravity at the one-loop level in Ref. \cite{Donoghue}.  

The infrared behavior of quantum gravity has been revisited recently in \cite{Naculich:2011ry,White:2011yy}
in the context of exploiting analogies between gauge theories to gravity \cite{Bern:2006kd}.     The study of perturbative quantum gravity amplitudes and cross sections also complements studies of nonperturbative quantum gravity at very high energies \cite{Giddings}.   In this paper we analyze amplitudes for fixed-angle scattering in quantum gravity.   We will identify at arbitrary orders the classes of diagrams that give collinear or soft infrared divergences, and generalize the cancellation of the former to energetic lines, for which the eikonal approximation does not apply in general.

We begin with a study of the collinear sector of quantum gravity, including its coupling to gauge theory matter, with the aim of complementing the work in Refs.\ \cite{Naculich:2011ry} and \cite{White:2011yy}, which concentrated primarily on soft gravitons.  We consider amplitudes with all massless external lines, all at fixed angles relative to each other, both incoming and outgoing.   We will show that for such ``wide-angle" scattering,  perturbative amplitudes are free of collinear singularities altogether to any fixed order, despite their presence on a diagram-by-diagram basis.  This result generalizes the observation of Ref.\ \cite{Weinberg}  in the eikonal approximation.  We go on to investigate soft graviton singularities, and conclude that they originate only from ladder exchange between finite-energy lines  \cite{Naculich:2011ry,White:2011yy}. To demonstrate these results, we will use the general infrared analysis developed for gauge theories in \cite{Sterman1}, \cite{Akhoury} and \cite{Libby:1978qf} and elaborated in \cite{Sterman:1994ce}.    To be specific, we consider the harmonic, or de Donder gauge for the quantization of quantum gravity, with perturbation theory rules, including ghosts \cite{Feynman:1963ax,'tHooft:1974bx} as  summarized, for example, in \cite{Capper:1973pv,Berends} and \cite{Donoghue:1994dn}.   Happily, we will not need the detailed features of the rules, only their covariance and a  counting of numbers of derivatives.

The method of \cite{Sterman1} begins with the observation that a necessary condition for infrared enhancement, whether soft or collinear, is the presence of pinch singularities in subspaces (pinch surfaces) of virtual loop momentum integrals \cite{Landau:1959fi,Edenetal}.   Each such pinch surface is conveniently characterized by a reduced diagram, consisting of the lines that are forced on-shell at the surface in question.    

\begin{figure}
\includegraphics[width=2.5in]{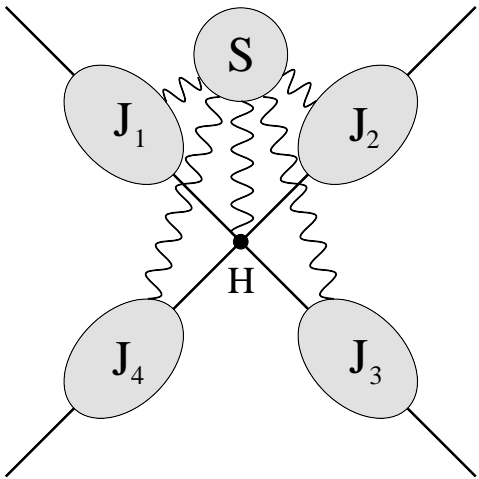}
\caption{A general reduced diagram at the pinched singular point. J and S denote the jet and soft subdiagrams and H is a hard vertex.   Here each line may represent any number of soft propagators connecting S to the remainder of the diagram, or the jets J${}_i$ to the hard part H.}
\label{Blob}
\end{figure}

This analysis is particularly straightforward for wide-angle scattering.   At leading power a single effective vertex in the reduced diagram mediates the hard scattering.   Specifically, an analysis of the pinch surfaces for wide-angle scattering gives for the most general reduced diagram involving only massless lines (including gravitons) the  form shown in Fig.\ \ref{Blob}, where for purposes of illustration, only four external legs are shown. The letters J and S denote respectively the jet and soft subdiagrams and H is a hard vertex \cite{Sterman1,Akhoury}.  
At the pinch surface, all lines in each Jet ${\rm J}_i$ are collinear to each other and to the external line $p_i$ to which they attach, all lines in S carry zero momentum, and all lines in H are off-shell.
In the following sections we study the nature of the various subdiagrams and find that a remarkably simple structure emerges, as suggested by the eikonal analysis of Ref.\ \cite{Weinberg}. The infrared singular behavior of quantum gravity is simpler than that of massless quantum electrodynamics.

In section \ref{sec:jet} we introduce a power-counting procedure to identify the types of reduced diagrams that yield collinear singularities in theories of  pure gravity and of gravity coupled to other matter. The infrared divergences in the pure matter sector have been studied extensively in the literature so we only focus on the new divergences that arise as a result of gravitational interactions. We find that the only types of diagrams that give mass divergences are those with no internal graviton jet loops, and which contain only three point vertices.  We also show that mass divergences do not arise in diagrams with both standard model particles and gravitons when only the gravitons attach to an external leg.

Next, in Sec.\ \ref{sec:soft} we develop a power-counting procedure to find what types of reduced diagrams with virtual gravitons give soft divergences. We find that for the case of virtual soft graviton corrections to a hard vertex, only diagrams of the ladder type give rise to soft divergences, as previously observed in Refs. \cite{Naculich:2011ry} and \cite{White:2011yy}.   In addition, we observe that the representation of divergent soft graviton interactions in terms of Wilson lines, as explicitly conjectured in \cite{Naculich:2011ry}, follows readily from the cancellation of collinear singularities.

Sec.\ \ref{sec:cancel} is devoted to the proof that  collinear singularities cancel when all collinear-divergent diagrams are combined, using a gravitational Ward identity. 
As an illustration of how the arguments of section \ref{sec:cancel} work, in the appendix we develop an extension of the analysis of Ref. \cite{Weinberg}, to show explicitly the cancellation of the graviton collinear singularities 
for kinematic regions where the eikonal approximation applies.

\section{Jet Power Counting}
\label{sec:jet}
\begin{figure}
\includegraphics[width=2.5in]{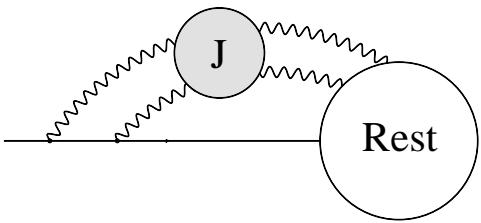}
\caption{A diagram with a graviton jet attached to an external leg. We show the case where there are two gravitons being emitted from the external line and two gravitons attaching to ``rest" but these can be any  numbers. The analysis in section \ref{sec:jet} shows that the number of gravitons emitted and attached do not need to be the same.}
\label{BaseFig}
\end{figure}

In this section we will build upon a power counting procedure for infrared divergences developed in  Refs. \cite{Sterman1, Akhoury, Libby:1978qf}.
Let us consider a graviton jet attached to a massless on-shell line with momentum $p$ in an arbitrary diagram as in Fig.\ \ref{BaseFig}. The graviton jet reattaches to the rest of the diagram, labeled ``rest" in the figure. For purposes of classification, gravitons that reattach to the same external leg from which they were emitted 
are considered as part of the jet, rather than attached
to ``rest".   Such external leg corrections do not give rise to collinear singularities, as we will show later.

Let $L_{J}$ be the number of loops, $N_{J}$ the number of lines in jet J, and $N_{num}$ the total power of what we will call ``normal variables" in the numerator that arise from the vertices and propagators of the jet. Normal variables   are chosen such that they vanish at the pinch singular point that causes the infrared divergence.  Then, singularities of the integrand appear through their dependence on normal variables.

The loop momentum integrand corresponding to any Feynman diagram can be made a homogenous function of the normal variables by keeping only the lowest power in both the numerator and denominator factors.  Counting powers of normal variables then enables us to determine the finiteness or potential for divergence of the pinch surface in question.   This is measured by the degree of divergence, given by the number of normal variables, minus the homogeneity (power in normal variables) of the product of denominators, plus the homogeneity for the numerators. For examples, see Eqs.\ \eqref{base} and \eqref{soft1} below.

In order to identify the normal variables for collinear singularities, let us make a change of variables in each jet loop integral such that:
\begin{align}
\int d^{4}l\sim\int dl_{\bot}^{2}dl^{+}dl^{-}\, ,
\end{align}
where $l_{\bot}^{2}$ includes the two components of the loop momenta $l$ that are transverse to $p$ and $l^{\pm}$ is defined as $\frac{1}{\sqrt{2}}(l_{0}\pm \vec{l}\cdot\hat{p})$, with $\hat p$ a unit 3-vector in the direction of the jet. This change of variables actually requires one to evaluate a Jacobian, but we omit this step as this factor will not contain any singularities. Note that for a collinear line we can rotate to a frame such that $l^{-}$ and $l_{\bot}^{2}/l^+$ become small, so we choose these as the normal variables (for convenience, both with dimensions of mass).  With this choice, each jet loop will contribute two normal variables to the total collinear degree of divergence of the diagram.  The consistency of this choice is discussed in Ref.\ \cite{Sterman:1994ce} and \cite{Collinsbook}.   A similar power counting plays a role in soft-collinear effective theory \cite{Bauer:2000yr}.

In the de Donder gauge, the graviton propagator is \cite{Berends}
\begin{align}
\frac{i}{2}[\eta^{\alpha\gamma}\eta^{\beta\delta}+\eta^{\beta\gamma}\eta^{\alpha\delta}-\eta^{\alpha\beta}\eta^{\gamma\delta}]\frac{1}{l^{2}+i\epsilon}\, .
\end{align}
Since $l^{2}=2l^{+}l^{-}-l_{\bot}^{2}$, each graviton propagator will be linear in normal variables in the denominator. Thus, each graviton jet line will contribute $-1$ to the degree of divergence from the diagram. In summary, we can write the total collinear degree of divergence of diagrams of the type shown in figure \ref{BaseFig} as
\begin{align}
\gamma_{CO}=2L_{J}-N_{J}+N_{num}\, .
\label{base}
\end{align}
A diagram can have a collinear singularity whenever $\gamma_{CO}\le0$, and $\gamma_{CO}=0$ corresponds to a logarithmic divergence.

As each jet line connects two vertices, we can also use the relation,
\begin{align}
2N_{J}=\sum_{i\ge3}iV_{i}+N_{G}\, ,
\label{lines}
\end{align}
where $V_{i}$ is the total number of $i$-point vertices in the diagram, not counting those that attach to ``rest" (see Fig.\ \ref{BaseFig}) and $N_{G}$ is the number of gravitons in the jet that attach to ``rest".  Here we treat ``rest" as an $N_G+1$-point vertex in the jet subdiagram, and we do not count the external line of the jet in $N_J$.

We can write the minimum homogeneity of normal variables in the numerator as
\begin{align}
N_{num}=\frac{1}{2}N_{mom}-\min[\frac{1}{2}N_{mom},\,\sum_{k}\frac{k}{2}N_{k}]\, ,
\label{numerator}
\end{align}
where $N_{mom}$ is the total power of momentum vectors from the vertices in the jet subdiagram and $N_{k}$ is the number of vertices with $k$ powers of momentum that are contracted with non-collinear vectors from ``rest". The factor of $\frac{1}{2}$ reflects that the scalar product of the momenta of any two jet lines is linear in the normal variables of the jet, and hence adds unity to the overall homogeneity of the numerator.  We subtract $\sum_{k}\frac{k}{2}N_{k}$ from $N_{num}$ because each graviton line attaches to ``rest" at a vertex involving some integer, $k$, of powers of momenta, collectively denoted $p'^{\mu}$, that are not collinear with $p^{\mu}$.  This is why we do not classify gravitons reattaching to the same leg they were emitted from as part of ``rest", as in this case $p'=p$.   The graviton propagator can then contract the momenta in the vertex in ``rest" with momentum vectors from a vertex in the graviton jet. This will result in terms of order $(p\cdot p')^{k}$. These terms are zeroth order in normal variables. That is, at each such vertex, $k/2$ factors of momenta that are nearly proportional to $p^\mu$ can ``escape" the jet, forming ``large" invariants that do not vanish at the pinch surface.  Thus, we must subtract ${k}/{2}$ from $N_{num}$ for each of the $N_{k}$ vertices from ``rest". The last term in (\ref{numerator}) is inserted simply to ensure that $N_{num}\ge0$ in all cases.

For gravity we make the expansion $g_{\mu\nu}=\eta_{\mu\nu}+\kappa h_{\mu\nu}$ and take the quantum field to be $h_{\mu\nu}$. Gravitational vertices correspond to terms in the Lagrangian that are symbolically of the form $\kappa^{i+j-4}\partial^{j}h^{i}$. Let $V_{i,j}$ be the number of $i$-point vertices with $j$ powers of momentum in the diagram (not including those in  ``rest"). This allows us to write $N_{mom}$ as
\begin{align}
N_{mom}=\sum_{j}j\sum_{i\ge3}V_{i,j}\, .
\label{vectors}
\end{align}
For the case of pure gravity we can write the Einstein-Hilbert Lagrangian as \cite{Goldberg}
\begin{align}
\mathcal{L}_{EH}&=\sqrt{-g}R=\frac{1}{2}(\hat{g}^{\alpha\kappa}\hat{g}_{\rho\sigma}\hat{g}^{\rho\sigma}_{,\alpha}+2\hat{g}^{\alpha\kappa}_{,\alpha})_{,\kappa}
\nonumber \\
&+\frac{1}{8}\hat{g}^{\alpha\kappa}_{,\rho}\hat{g}^{\lambda\beta}_{,\sigma}(2\hat{g}^{\rho\sigma}\hat{g}_{\lambda\alpha}\hat{g}_{\kappa\beta}-\hat{g}^{\rho\sigma}\hat{g}_{\alpha\kappa}\hat{g}_{\lambda\beta}-4\eta^{\sigma}_{\kappa}\eta^{\rho}_{\lambda}\hat{g}_{\alpha\beta})\, ,
\end{align}
where a comma denotes an ordinary partial derivative and $\hat{g}^{\alpha\beta}=\sqrt{g}g^{\alpha\beta}$.
When written in this form, it is easy to see that $\mathcal{L}_{EH}$ only has terms with $j=2$, so we take all $i$-point graviton vertices as having two powers of momentum. Of course from an effective field theory point of view  \cite{Donoghue:1994dn} there should be higher order terms in $R$ but this will serve only to increase $j$ in the vertices. Since it is clear from equations \eqref{base}, \eqref{numerator}, and \eqref{vectors} that increasing the number of derivatives, $j$ will only increase $\gamma_{CO}$, considering only the linear term in $R$ gives the most infrared divergent case. 

The coupling of matter to gravity is given by $\kappa h_{\mu\nu}T^{\mu\nu}$, where $T^{\mu\nu}$ is the energy momentum tensor of the matter field. For bosons, the energy momentum tensor has at least two derivatives, so this situation is similar to the pure gravity case. For fermions, the energy momentum tensor is proportional to 
$\bar{\psi}\gamma^{\mu}(\overrightarrow{\partial}^{\nu} - \overleftarrow{\partial}^{\nu})\psi$. However, for power counting purposes, using the Gordon identity we may replace $\gamma^{\nu}$ by $\partial^{\nu}/m$, where $m$ is the fermion mass.    (For massless fermions, the vanishing normalization of spinors leads to simply $\partial^\nu$.)

If we combine the Euler identity (note that for the jet subdiagram we consider ``rest" to be a vertex so the number of vertices in the jet subdiagram is $\sum_{i}V_{i}+1$),
\begin{align}
L_{J}=N_{J}-\sum_{i\ge3}V_{i}\, ,
\end{align}
with equations \eqref{base}, \eqref{lines}, \eqref{numerator}, and \eqref{vectors} we get the relation,
\begin{align}
\gamma_{CO} & = \sum_{i\ge3}(\frac{i}{2}-2)V_{i}+\frac{1}{2}N_{G}+\frac{1}{2}\sum_{j}j\sum_{i\ge3}V_{i,j}
\nonumber \\
&-\min[\frac{1}{2}\sum_{j}j\sum_{i\ge3}V_{i,j},\,\sum_{k}\frac{k}{2}N_{k}]\, .
\end{align}

Let us consider the case where
\begin{align}
\min[\frac{1}{2}\sum_{j}j\sum_{i\ge3}V_{i,j},\,\sum_{k}\frac{k}{2}N_{k}]=\sum_{k}\frac{k}{2}N_{k}\, ,
\end{align}
as this is the more divergent case (since the other option is available solely to prevent $N_{num}$ having an artificially negative contribution to $\gamma_{CO}$). In this case,
\begin{align}
\gamma_{CO} = \sum_{i\ge3}(\frac{i}{2}-2)V_{i}+\frac{1}{2}N_{G}+\frac{1}{2}\sum_{j}j\sum_{i\ge3}V_{i,j}-\sum_{k}\frac{k}{2}N_{k}\, .
\label{power}
\end{align}
We readily deduce the consequences of this result.

\subsubsection*{Ladder Diagrams}
\begin{figure}
\includegraphics[width=2.5in]{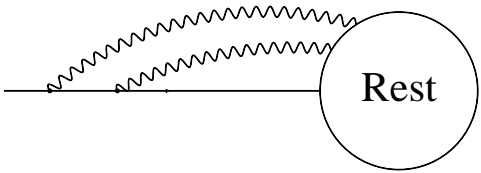}
\caption{A typical ladder diagram.}
\label{Ladder}
\end{figure}

As a first example, let us consider the simplest case of ladder diagrams (Fig.\ \ref{Ladder}). In this case, we only have 3-point vertices, which have two powers of momentum, and no internal jet loops. Thus $i=3$, and $j=2$. For the case of gravitational couplings $k=2$, so $\sum_{k}\frac{k}{2}N_{k}=N_{G}$. Applying these conditions we have
\begin{align}
\gamma_{CO}=-\frac{1}{2}V_{3}+V_{3,2}-\frac{1}{2}N_{G}\, .
\end{align}
For ladder diagrams it is easy to see that $V_{3}=V_{3,2}=N_{G}$, so we have $\gamma_{CO}=0$, which corresponds to logarithmic divergence.

\subsubsection*{Diagrams with only three point vertices}
\begin{figure}
\includegraphics[width=2.5in]{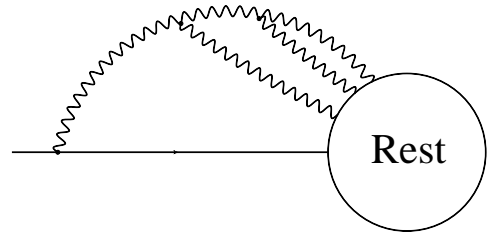}
\caption{A diagram with no internal jet loops and only three point vertices.}
\label{Tree}
\end{figure}

Once again we have $i=3$, $j=2$. So, $V_{3}=V_{3,2}$ and again for gravitational couplings $\sum_{k}\frac{k}{2}N_{k}=N_{G}$. Therefore we (again) have
\begin{align}
\gamma_{CO}=\frac{1}{2}(V_{3}-N_{G})\, .
\end{align}
Note in the case of diagrams with no internal jet loops as in Fig.\ \ref{Tree}, we have $V_{3}=N_{G}$ and again we have a logarithmic collinear divergence. On the other hand, if we add any internal jet loops $V_{3} > N_{G}$ and there is no collinear singularity. Adding a four-point (or higher) vertex will only increase the collinear degree of divergence and will prevent a mass divergence. Thus, the only diagrams that give mass divergences are those with no internal jet loops, and with only three-point vertices. These include the ladder diagrams discussed as the first example. 

\subsubsection*{External Leg Corrections}
\begin{figure}
\includegraphics[width=2.5in]{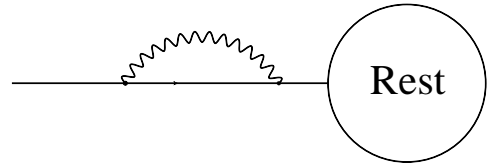}
\caption{An example of an external leg correction. These do not result in collinear singularities.}
\label{TwoPoint}
\end{figure}

As mentioned earlier, graviton lines that reattach to the same leg from which they were emitted are not considered as lines that attach onto ``rest". This is because in this case the momenta at the two vertices the graviton line connects are collinear, so we get a numerator factor that is quadratic in normal variables. Thus, these vertices do not contribute to the subtraction of $N_{k}$ from momentum factors in the numerator in equation \eqref{numerator}. Because of this, diagrams such as the one in Fig.\ \ref{TwoPoint} do not have collinear singularities. For Fig.\ \ref{TwoPoint} in particular, using Eq.\ \eqref{power} we see that $\gamma_{CO}=1$. Gauge invariance ensures the cancellation of the single particle pole (which for power counting purposes corresponds to matching  $+1$ from the normal variables of the numerator with  $-1$ from the on-shell propagator.)
We can see from equation \eqref{power} that adding further graviton lines cannot decrease the collinear degree of divergence wherever they are attached. Thus, self energy and other diagrams with graviton lines that reattach to the same leg from which they were emitted do not have collinear singularities.

\subsubsection*{Diagrams with gravitons turning into standard model particles}
\begin{figure}
\includegraphics[width=2.5in]{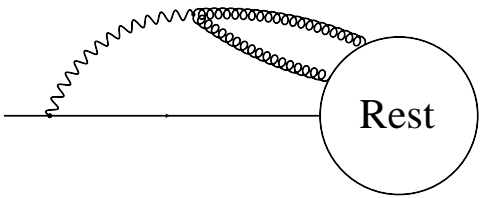}
\caption{A graviton jet turning into two gluons.}
\label{Gluon}
\end{figure}

\begin{figure}
\includegraphics[width=2.5in]{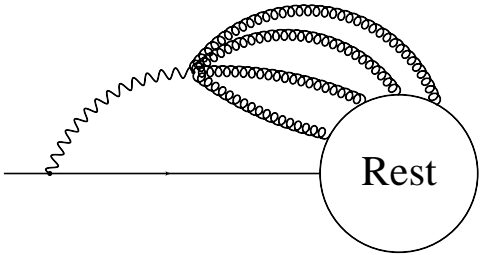}
\caption{A graviton jet turning into four gluons.}
\label{FourGluon}
\end{figure}

Let us consider the case where a graviton emitted from one of the external line turns into standard model particles such as gluons. Such an example is given in Fig.\ \ref{Gluon}. For this particular case, Eq.\ \eqref{base} tells us that
\begin{align}
\gamma_{CO}=2(2)-4(1)+N_{num}=N_{num}\, .
\end{align}
If the gluons were gravitons, according to Eqs.\ \eqref{numerator} and \eqref{vectors}, we would have $N_{num}=\frac{1}{2}(4)-2=0=\gamma_{CO}$ and thus a logarithmic collinear singularity. This is because $k=2$ for all vertices of ``rest" to which gravitons attach. However, for the case of gluons we would have $k=1$ for all vertices on ``rest". This means that for the case of gluons, $N_{num}=\frac{1}{2}(4)-1=1=\gamma_{CO}$, so there is no collinear singularity. Note that since $\gamma_{CO}\ge0$ for any diagram involving just gravitons, adding further graviton lines does not change the situation. Adding a higher point gluon-graviton vertex such as the one in Fig.\ \ref{FourGluon} does not help, as in this case the the contribution to $\gamma_{CO}$ from $L_{J}$ and $N_{J}$ will already be positive and $N_{num}$ is at least zero. For instance, for the process shown in Fig.\ \ref{FourGluon}, $2L_{J}-N_{J}=2$. Thus, it is impossible to have a collinear-divergent diagram with both standard model particles and gravitons where only the gravitons attach to the on-shell line.
Precisely the same reasoning applies to the vector ghosts of quantum gravity \cite{Feynman:1963ax},
because, although their interactions with gravitons are not identical to those of
photons or gluons, the numbers of derivatives at the vertex is the same.

In summary, we have found that collinear divergences may be found in diagrams: (1) with no internal jet loops, (2) with only three-point 
vertices among gravitons, and (3) that do not link gravitons to collinear standard model particles.   We now turn to soft divergences.

\section{Soft Power Counting}
\label{sec:soft}

So far, we have concentrated on the ``jet" subdiagrams of the arbitrary pinch surface represented in Fig.\ \ref{Blob}.  We now describe the inclusion of 
interacting soft gravitons \cite{Naculich:2011ry,White:2011yy}, and show how our power counting arguments confirm the conclusion that to fixed order only ladder-like graphs show soft divergence, factorizable onto products of Wilson lines.

\begin{figure}
\begin{center}
\includegraphics[width=2.5in]{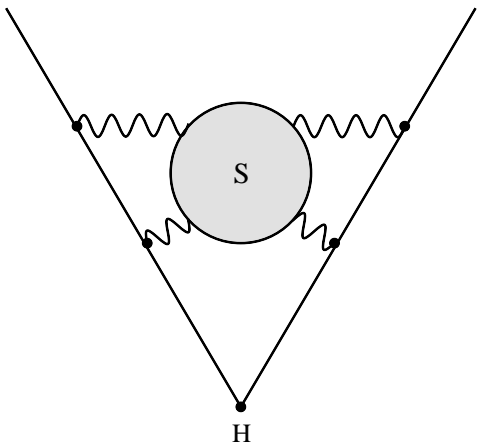}
\end{center}
\caption{Arbitrary diagram with virtual soft graviton corrections to a hard vertex. The finite momentum lines are drawn with a solid line.}
\label{BaseSoft}
\end{figure}

\begin{figure}
\begin{center}
\includegraphics[width=2.5in]{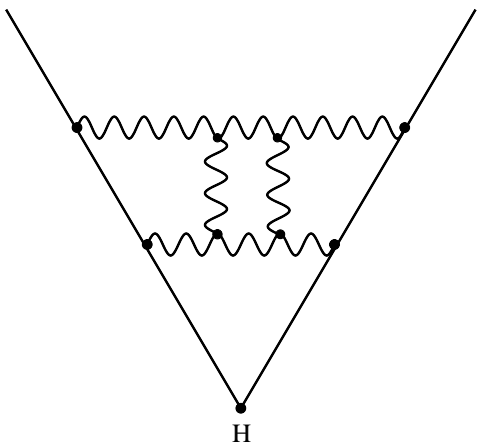}
\end{center}
\caption{An example of the type of diagram that we discuss in section \ref{sec:soft}. In this diagram $L_{S}=4$, $N_{S}=8$, and $N_{E}$=4.}
\label{SoftEx}
\end{figure}

We can carry out power counting for soft divergences simultaneously with collinear divergences, by considering diagrams of the type shown in Fig.\ \ref{BaseSoft}, which shows soft virtual graviton corrections to a hard vertex. All of the graviton lines in S are soft. Each solid line represents any set of collinear jet lines.  In principle, the particles represented by the solid finite momentum lines can have any spin, as the dominant coupling of soft gravitons is to momentum flow, and independent of spin \cite{Weinberg}.  We now define, by analogy to Eq.\ (\ref{base}) for the degree of collinear divergence, a degree of soft divergence,
\begin{align}
\gamma_{soft}=4L_{S}-2N_{S}-N_{E}+N_{sn}\, ,
\label{soft1}
\end{align}
where $L_{S}$ is the number of loops in S, including loops that link S with the jets, $N_{S}$ is the number of soft graviton lines, $N_{E}$ is the number of virtual finite momentum lines in the diagram, and $N_{sn}$ is the contribution of soft normal variables to the numerator (for clarification of these quantities, see the example given in Fig.\ \ref{SoftEx}).  Specifically, we may take $N_E$ to denote the {\it change} in the number of finite-momentum lines due to the attachment of soft gravitons to the jet.   For simplicity, therefore, we choose the most singular case, in which all soft gravitons attach to the jets at vertices with only two finite-momentum lines.   

For the soft subdiagram  we can choose the normal variables to be all four components of the loop momenta, so that there is a factor of four times $L_{S}$ in \eqref{soft1}.  With this choice, all soft graviton denominators are quadratic in normal variables, so there is a factor of minus two multiplying $N_{S}$.  The denominators of the propagators corresponding to the (nearly on-shell) virtual finite momentum lines are, by contrast, linear in graviton momenta.   Thus, $N_{E}$ is associated with a factor of minus one.   
Notice that this requires the scales of normal variables for soft and collinear momenta to be the same.   If, for example, the soft normal variables are larger than the collinear normal variables, the denominators of finite energy lines will be independent of the latter, which would eliminate collinear singularities.

The linearity of finite energy lines in soft normal variables, combined with the dominance of jet momenta in the coupling of soft gravitons to finite energy lines is equivalent to the eikonal approximation.   We note that for wide angle scattering there are no additional pinches that would invalidate the eikonal approximation.   We will return to this point below.

\begin{figure}
\begin{center}
\includegraphics[width=2.5in]{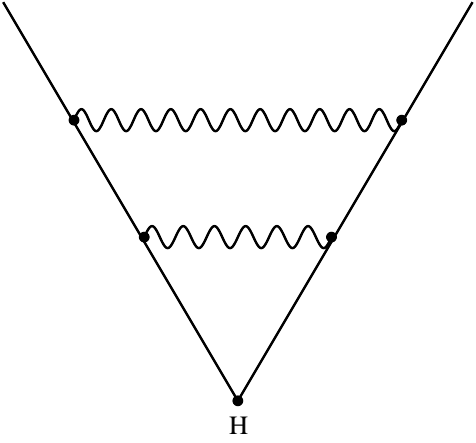}
\end{center}
\caption{An example of a diagram with only three point couplings to hard lines that will lead to a soft divergence.}
\label{Soft}
\end{figure}

Using similar reasoning as with the case of collinear power counting we arrive at the relations,
\begin{align}
&2N_{S}=\sum_{i\ge3}iV_{i}+\sum_{m}mN^{(m)}_{hs}\, ,
\label{soft2}
\\
&N_{E}=N_{hs}\, ,
\label{soft3}
\\
&N_{sn}=\sum_{j}j\sum_{i\ge3}V_{i,j}\, ,
\label{soft4}
\\
&L_{S}=N_{S}+N_{E}-\sum_{i\ge3}V_{i}-N_{hs}\, ,
\label{soft5}
\end{align}
where again $V_{i}$ is the number of $i$-point vertices in S and $V_{i,j}$ are the number $i$-point vertices in S with $j$ powers of momentum. The term $N^{(m)}_{hs}$ is the number of vertices at which $m$ soft lines attach to a finite momentum line, and $N_{hs}=\sum_{m}N^{(m)}_{hs}$. Note the vertices that count towards $N_{hs}$ do not contribute to the homogeneity of normal variables from the numerator, $N_{sn}$, as their numerator momentum factors are given to leading power by the momenta of finite momentum lines, independent of soft normal variables. Combining equations \eqref{soft1}, \eqref{soft2}, \eqref{soft3}, \eqref{soft4}, and \eqref{soft5}, we get
\begin{align}
\gamma_{soft}=\sum_{i\ge3}(i-4)V_{i}+\sum_{j}j\sum_{i\ge3}V_{i,j}+\sum_{m}mN^{(m)}_{hs}-N_{hs}\, .
\label{softfinal}
\end{align}
Using this relation, we see that if there are no soft graviton vertices in S ($V_{i}=0$, $V_{i,j}=0$) and only vertices with one soft graviton attached to the finite momentum lines ($\sum_{m}mN^{(m)}_{hs}=N^{(1)}_{hs}=N_{hs}$) then $\gamma_{soft}=0$, indicating a logarithmic soft divergence. An example of a diagram that gives a logarithmic divergence is shown in Fig.\ \ref{Soft}. On the other hand if there is even one soft vertex in S with $j\ge2$ (as is the case for graviton vertices), or if there is even one vertex with more than one soft graviton coming off a finite momentum line (in which case $\sum_{m}mN^{(m)}_{hs}>N_{hs}$) then $\gamma_{soft}>0$ and there is no soft divergence. Thus, in agreement with \cite{Naculich:2011ry,White:2011yy} we have seen that the only diagrams that give rise to soft graviton divergences are ladder and crossed ladders with only three point vertices where the ladders attach to finite momentum lines.

The above conclusions apply whether or not the jet subdiagrams consist of single lines or contain loops.   As we shall see in the next section, however, collinear singularities associated with nontrivial jet subdiagrams cancel, leaving only single finite-energy lines to couple to the soft gravitons.   In this sense, the factorization of soft infrared gravitons conjectured in Ref.\ \cite{Naculich:2011ry} is automatic, because as noted above, soft divergences are reproduced by considering only the linear dependence of finite-energy denominators on soft graviton momenta.  In addition, as we have seen, infrared divergences are associated with the coupling of soft gravitons to a finite-energy graviton (or matter) line of momentum $p$ through the vertex $\kappa p^\mu p^\nu$ only.  Together, these features of soft graviton infrared divergences are precisely the perturbation theory rules of the Wilson lines described in Ref.\ \cite{Naculich:2011ry}, and this discussion serves as a proof of the conjecture there.   The importance of the cancellation of collinear singularities was also noted in Ref.\ \cite{White:2011yy}, and we now turn to a proof of this cancellation independent of the eikonal approximation for the collinear gravitons.

\section{Cancellation of collinear singularities}
\label{sec:cancel}
In this section we give a general argument for the cancellation of gravitational collinear singularities using the basic gravitational Ward identity \cite{Weinberg2} that decouples unphysical graviton polarizations from physical processes.  Our argument is independent of the eikonal approximation.  A combinatoric proof of this cancellation along the lines of \cite{Weinberg}, in the special case of the eikonal approximation, is given in the appendix.

We have shown by power counting that collinear singularities require graviton jets that have no internal loops and only three point vertices. Let us consider the addition of such a jet to an arbitrary external line with momentum $p^{\mu}$ as in Fig.\ \ref{BaseFig}, where $j$ gravitons with momentum $l^{\mu}_{i}$ and the now off-shell continuation of the original external line with momentum $p^{\mu}-\sum l_{i}^{\mu}$ attach to ``rest". We can represent the matrix element corresponding to such a diagram as \cite{Collins:1989gx}
\begin{align}
i\mathcal{M}=\int\frac{d^{4}l_{i}}{(2\pi)^{4}}&J(p^{\mu}-\sum_{i=1}^j l_{i}^{\mu_i},\{  l_{i}^{\mu_i}\})_{\{\mu_{i}\nu_{i}\}}
\nonumber \\
\times&R(p^{\mu}-\sum_{i=1}^j l_{i}^{\mu},\{  l_{i}^{\mu_i}\})^{\{\mu_{i}\nu_{i}\}}\, ,
\label{factororig}
\end{align}
where $R$ and $J$ correspond to the ``rest"  and jet subdiagrams, respectively, and where $\{\mu_{i}\nu_{i}\}$ represent  the $2j$ spacetime indices corresponding to the $j$ collinear gravitons attached to ``rest".

\begin{figure}
\includegraphics[width=2.5in]{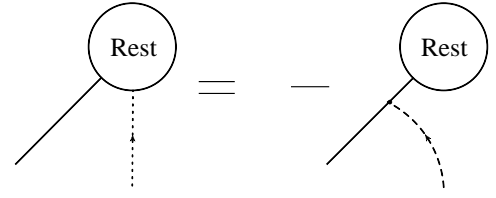}
\caption{The Ward identity for the case $n=1$}
\label{WardIdentity1}
\end{figure}

Since all the lines attaching to $R$ from $J$ have momenta collinear with the jet momentum $p^{\mu}$, we can isolate the leading power behavior near the collinear pinch surface, by making the replacement $l^{\mu} \rightarrow l_{\alpha}\bar{v}^{\alpha}v^{\mu}$, where we define the lightlike vectors $v^{\mu}=\delta^{\mu}_{+}$, and $\bar{v}^{\mu}=\delta^{\mu}_{-}$.  Here we are working in the basis of normal variables where momentum vectors now have the components $l^{\mu}=(l^{+},l^{-},l_{\perp})$ and $l^{2}=2l^{+}l^{-}-l_{\perp}^{2}$. After making this replacement and pulling out the factors of $v^{\mu}$ and $\bar{v}^{\mu}$, we can write the function $J$ as
\begin{align}
&J(p^{\mu}-\sum l_{i}^{\mu_i},\{  l_{i}^{\mu_i}\})_{\{\mu_{i}\nu_{i}\}}
\nonumber \\
\rightarrow & J(p^{\alpha}-\sum l_{i}^{\alpha_i},\{  l_{i}^{\alpha_i}\})_{\{\alpha_{i}\beta_{i}\}}\prod_{i}\bar{v}^{\alpha_{i}}\bar{v}^{\beta_{i}}v_{\mu_{i}}v_{\nu_{i}}\, .
\end{align}
It is easy to see that to leading power in normal variables
\begin{align}
J(p^{\alpha}-\sum l_{i}^{\alpha_i},\{ l_{i}^{\alpha_i}\})_{\{\alpha_{i}\beta_{i}\}}
\bar{v}^{\alpha_{i}}\bar{v}^{\beta_{i}}v_{\mu_{i}}v_{\nu_{i}} &=
 \nonumber\\
& \hspace{-60mm}
J(p^{\alpha}-\sum l_{i}^{\alpha_i},\{  l_{i}^{\alpha_i}\})_{\{\alpha_{i}\beta_{i}\}}
 \frac{n^{\alpha_{i}}}{n\cdot l_{i}}\frac{n^{\beta_{i}}}{n\cdot l_{i}}l_{i,\mu_{i}}l_{i,\nu_{i}}\, ,
\label{MomId}
\end{align}
for any vector $n^{\alpha}$ that is not collinear with $l^{\mu}_{i}$ by making the substitution $l^{\mu} \rightarrow l_{\alpha}\bar{v}^{\alpha}v^{\mu}$ on the right hand side of \eqref{MomId}. Thus, we can write \eqref{factororig} as
\begin{align}
i\mathcal{M}=\int\frac{d^{4}l_{i}}{(2\pi)^{4}}&J(p^{\alpha}-\sum l_{i}^{\alpha_i},\{  l_{i}^{\alpha_i}\})_{\{\alpha_{i}\beta_{i}\}}
\nonumber \\
\times &\prod_{i}\frac{n^{\alpha_{i}}}{n\cdot l_{i}}\frac{n^{\beta_{i}}}{n\cdot l_{i}}l_{\mu_{i}}l_{\nu_{i}}
\nonumber \\
\times&R(p^{\mu}-\sum l_{i}^{\mu_i},\{  l_{i}^{\mu_i}\})^{\{\mu_{i}\nu_{i}\}}\, .
\label{factorfinal}
\end{align}
This is known as the ``collinear approximation" \cite{Collins:1989gx,Collins:1982wa}. We see from \eqref{factorfinal} that in
the collinear approximation the $j$ gravitons that attach to ``rest" are longitudinally, or ``scalar", polarized. This allows us to use the on-shell Ward identity for gravitons \cite{Weinberg2}, which inforces the decoupling of such unphysical polarizations,
\begin{align}
l_{\mu_{j}}l_{\nu_{j}}\mathcal{S}^{...\mu_{j}\nu_{j}...}=0\, .
\label{WardEq}
\end{align}
where $\mathcal{S}^{...\mu_{j}\nu_{j}...}$ is an arbitrary S-matrix element with the polarization tensors factored out. For our particular case, we can apply the Ward identity to the subdiagram ``rest", which includes an external line with momentum $p^{\mu}-\sum l_{i}^{\mu}$. In R, the latter may be considered on-shell with physical or scalar polarization, up to corrections that are higher order in normal variables. Diagramatically this can be expressed by Fig.\ \ref{WardIdentity1}, which corresponds  a single longitudinally, or scalar, polarized graviton ($n=1$) attached to ``rest", which is drawn with a dotted line. The Ward identify \eqref{WardEq} tells us that the sum of the diagrams shown on the left hand side of Fig.\ \ref{WardIdentity1} and the attachment of the longitudinally polarized graviton on the external line shown on the right hand side is zero. The right hand side of Fig.\ \ref{WardIdentity1} has the same collinear degree of divergence as a self energy correction, which we previously showed does not have a collinear singularity.

\begin{figure}
\includegraphics[width=2.5in]{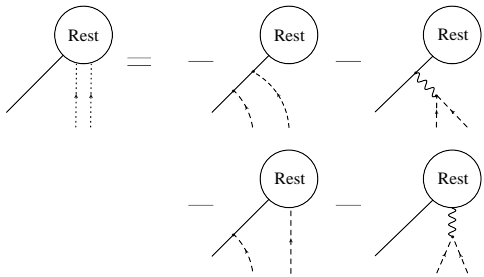}
\caption{The Ward identity for the case $n=2$. We do not explicitly show the diagrams where the longitudinal gravitons are interchanged.}
\label{WardIdentity2}
\end{figure}

The application of the Ward identity to the case of two longitudinally polarized gravitons ($j=2$) attached to ``rest" is shown in Fig.\ \ref{WardIdentity2}. Note the first and third diagrams on the right hand side of Fig.\ \ref{WardIdentity2} contain a self energy-like correction, so these are not collinearly divergent. The second and fourth diagrams result in an internal jet loop, and we know diagrams containing such loops are also not collinear-divergent. So there are no collinear singularities for this case either. Thus, the use of the Ward identities on \eqref{factorfinal} allows us to rule out any collinear singularity in the sum of diagrams contributing to it.   For gauge theories, by contrast, collinear singularities remain in the corresponding set of diagrams when the Ward identities are applied \cite{Collins:1989gx}.   The application of this reasoning to collinear gauge particles results in the factorization, rather than cancellation, of collinear singularities.

It is clear that adding any number of longitudinally polarized gravitons and applying the Ward identity in the manner above will always result in either additional self energy-like corrections or internal jet loops, which as we have seen preclude collinear singularities. Thus, while diagrams where an external leg emits a graviton jet with no internal loops and only three point vertices may be divergent on a diagram by diagram basis, when we consider the attachment of such a jet with the rest of the diagram, the Ward identity insures that collinear singularities cancel.  It is instructive to verify the cancellation for the case where collinear gravitons are relatively soft, so that we may apply the eikonal approximation. This is shown in the appendix.

\section{Conclusion}

In this paper we have introduced a power-counting procedure in order to see what types of reduced diagrams yield collinear and soft divergences in perturbative quantum gravity. For the fixed-angle elastic scattering amplitudes that we have considered, we find that the only types of reduced diagrams that give soft divergences are those of the ladder and crossed ladder type, where the soft gravitons interact only with finite momentum lines and not with each other \cite{Naculich:2011ry,White:2011yy}. These diagrams give rise to logarithmic soft divergences, which do not cancel when all diagrams of a given order are summed for an amplitude with fixed external lines.

For the case of collinear singularities, we see that the only types of diagrams that give mass divergences are those with no internal jet loops and only three-point vertices. These include the ladder and crossed ladder diagrams  \cite {Weinberg}. When all possible diagrams of this class are summed the Ward identity insures that the collinear singularities cancel. This is in contrast to the case of massless QCD or other massless gauge theories, where collinear singularities factorize rather than cancel.

The absence of collinear singularities has its basis in classical physics, where gravitational radiation is more suppressed in the collinear direction than electromagnetic radiation. Indeed, the leading multipole contributing to electromagnetic radiation is dipole while for gravitational radiation it is quadrupole. This can be seen conveniently in the rates of the lowest order modes of radiation in the two theories \cite{VanN}. For electromagnetic radiation sourced by an electric dipole with dipole moment $p$ oscillating with frequency $\omega$, the energy rate is
\begin{align}
\frac{d^{2}E}{d\Omega dt}=\omega^{2}(\frac{\omega^{2}p^{2}}{8\pi c^{3}})\sin^{2}\theta\,.
\end{align}
For gravitational radiation sourced by a mass M with trajectory ${\bf R}(t)={\bf R}_{0}\sin\omega t$ the energy rate is
\begin{align}
\frac{d^{2}E}{d\Omega dt}=\frac{G\omega^{6}M^{4}R_{0}^{4}}{4\pi c^{3}}\sin^{4}\theta,
\end{align}
which clearly shows the additional suppression in the forward direction.

The discussion in this paper has dealt with wide-angle scattering only.
These methods may be useful as well, however,
 in studies of higher order corrections for the Regge limit in quantum
gravity \cite{Giddings:2010pp}, and of higher-order cancellations between
virtual and real radiation.

\acknowledgements
R.A. and R.S. are supported by a grant from the U.S. Department of Energy.
The work of G.S.\ was supported by the
National Science Foundation, 
grant PHY-0969739.

\appendix
\section{Cancellation of Collinear Singularities in the Eikonal Approximation}
\label{sec:AB}

In this appendix, we review how the 
cancellation of collinear singularities is realized in the eikonal approximation, as discussed
by Weinberg in Ref.\ \cite{Weinberg}.   We present this argument for completeness, and also because
it confirms the use of the Ward identity illustrated by Figs.\ \ref{WardIdentity1} and \ref{WardIdentity2}.    
In particular, we emphasize that although we derive the factorization
(and hence cancellation) of collinear
gravitons from the hard scattering by using the Ward identity of Eq.\ (\ref{WardEq}), at no
point does the hard scattering, whether in the eikonal approximation or not, 
include all the diagrams of an S-matrix element (see Fig.\ \ref{WardIdentity2}).

We start by rewriting Eq.\ (\ref{factorfinal}), using the left-hand side of
Eq.\ (\ref{MomId}),
\begin{eqnarray}
i\mathcal{M}&=&\int\frac{d^{4}l_{i}}{(2\pi)^{4}}
J_m(p_{m}^{\alpha}-\sum l_{i}^{\alpha_i},\{  l_{i}^{\alpha_i}\})_{\{\alpha_{i}\beta_{i}\}}
\nonumber \\
&\ & \hspace{5mm}\times \prod_{i=1}^j\bar{v}_m^{\alpha_{i}}\bar{v}_m^{\beta_{i}}v_{m\hspace{0.1mm}\mu_{i}}v_{m\hspace{0.1mm}\nu_{i}}
\nonumber \\
&\ & \hspace{5mm}
\times R^{\rm (eik)}(p_{m}^{\mu}-\sum l_{i}^{\mu_i},\{  l_{i}^{\mu_i}\},\{p_{n}\})^{\{\mu_{i}\nu_{i}\}}\, ,
\nonumber\\
\label{factorfinal2}
\end{eqnarray}
where we now consider explicitly all of the $E$ external legs in the full diagram. 
The indices $i$ refer to gluons $l_i$ attaching $J_m$ to $R$, $i=1\dots j$.
The external leg connected to the jet we consider has momentum $p_{m}$ (the nearly on-shell portion of this line is not included in $R$) and the remaining $E-1$ non-collinear external legs have momentum $p_{n}$, $n \ne m$ (these non-collinear legs are included explicitly in $R$).

In this discussion, we follow Ref.\ \cite{Weinberg} by taking for the function $R$, representing
the remainder of the diagram,
a product of eikonal lines, linked at a point-like vertex $H(\left\{p_n\}\right)$,
and treat all connections of collinear gravitons to $R$ in the eikonal approximation,
summing over all diagrams.

Let us denote by $P\left(\{N_n\}\right)$
any {\it unordered} partition of the $j$ external gravition lines of $J_m$ into a set of bins 
with $N_n$ gravitons attached to line $p_n$.   At fixed momenta $l_i$, 
each ordering corresponds to a distinct diagram, and we must still
sum over all orderings of graviton connections to each
line $p_n$, $n\ne m$.   In these terms we write the contraction of the function $R$ with vectors $v$ in Eq.\ \eqref{factorfinal2} as
\begin{eqnarray}
R^{\rm (eik)}(p_m^{\mu}-\sum l_{i}^{\mu_i},\{  l_{i}^{\mu_i}\},\{p_{n}\})^{\{\mu_{i}\nu_{i}\}}\, \prod_{i}v_{m\hspace{0.1mm}\mu_{i}}v_{m\hspace{0.1mm}\nu_{i}}
&=& \nonumber\\
&\ & \hspace{-80mm}
\sum_{P(\{N_n\})}
{\cal E}_{\rm num}(\{N_n\})\, {\cal E}_{\rm den}(\{N_n\})\, H(\left\{p_n\},(\{N_n\})\right)\, .
\nonumber\\
\label{Reikdef}
\end{eqnarray}
In this expression, the eikonal numerator factors are given by
\begin{eqnarray}
{\cal E}_{\rm num}(\{N_n\})
=
\prod_{n \ne m}^{E-1}((p_{n}\cdot v_{m})^{2})^{N_{n}}\, ,
\label{calEnum}
\end{eqnarray}
since each graviton attached to a non-collinear external line contributes a numerator factor of $(p_{n}\cdot v_{m})^{2}$.

Similarly,
${\cal E}_{\rm den}$ summarizes all eikonal denominators, 
including the sum over 
orderings of graviton lines (labeled with the index $i$) from the jet $J_m$
to each of the other incoming massless lines, $p_n$
for a given choice of $P\left(\{N_n\}\right)$.    That is, for each partition $P$, we sum over
all permutations $\pi(N_n)$ of the connections of these lines 
to each of the $p_n$.   To the sum of these connections we may apply for each
external line $p_n$ the well-known identity for eikonal denominators, giving
\begin{eqnarray}
{\cal E}_{\rm den}(\{N_n\}) &=& 
\prod_{n\ne m}\, \sum_{\pi(N_n)}\, \prod_{a=1}^{N_n} 
\left( \sum_{i=1}^a  p_n\cdot q^{(n)}_{\pi_{N_n}(i)} \right)^{-1}
\nonumber\\
&=&
\prod_{n\ne m}\, \prod_{i=1}^{N_n} (p_n\cdot q_{i}^{(n)} )^{-1}\, ,
\label{calEdenom}
\end{eqnarray} 
where $q_{i}^{(n)}$ denotes the momentum of the $i$th graviton attached to the $n$th non-collinear external line. The subscript $\pi_{N_n}$ denotes that it is the momenta $q_{i}^{(n)}$ that we are permuting over.
This identity shows that the momentum dependence associated with denominators factors
into simple products for each external graviton of $R^{\rm (eik)}$.

Before combining equations \eqref{calEnum} and \eqref{calEdenom} 
for the eikonal numerator and denominator factors,
respectively we define 
\begin{eqnarray}
q_{i}^{(n)}=\alpha_{i}v_{m}\, ,
\end{eqnarray}
which holds in the leading collinear region. 
Each momentum $q_i^{(n)}$ and hence each $\alpha_i$, is
independent of to which $p_n$ the collinear graviton attaches.
In these terms, we have
 \begin{eqnarray}
{\cal E}_{\rm den}(\{N_n\}) &=& 
\prod_{{\rm all}\ i}\,\prod_{n\ne m} \frac{1}{\alpha_{i}}\  \left( p_n\cdot v_m \right)^{-N_n}\, ,
\label{calEdenom2}
\end{eqnarray} 
 so that the energy-dependence of the collinear gravitons
 is collected into a universal factor
 that is independent of the partition $P$.    Substituting 
 the numerator and denominator forms (\ref{calEnum}) and (\ref{calEdenom2})
into Eq.\ (\ref{Reikdef}) for $R^{\rm (eik)}$, we find
\begin{eqnarray}
R^{\rm (eik)}(p_m^{\mu}-\sum l_{i}^{\mu},\{  l_{i}^{\mu}\},\{p_{n}\})^{\{\mu_{i}\nu_{i}\}}\, \prod_{i}v_{m\hspace{0.1mm}\mu_{i}}v_{m\hspace{0.1mm}\nu_{i}}
&=&  
\nonumber\\
&\ & \hspace{-65mm}
 H(\left\{p_n\},(\{N_n\})\right)\,  \left( \prod_{{\rm all}\ i} \frac{1}{\alpha_{i}}\right)
\nonumber\\
&\ & \hspace{-65mm} \times\ \sum_{P(\{N_n\})}\,
\left(\prod_{n \ne m}^{E-1}(p_{n}\cdot v_{m})^{N_{n}}\right)
\, .
\label{Reik2}
\end{eqnarray}
In this form we see explicitly that the factor $R^{\rm (eik)}$ depends only on the 
numbers $N_n$ of each
unordered assignment of the collinear gluons $l_i$ to the external lines $p_n$.
For fixed $\{N_n\}$, the result is the same for every choice
of unordered partition $P\left(\{N_n\}\right)$.    

We may make this independence
explicit by replacing the sum over unordered assignments 
by a sum over all $N_n$ that add up to $j$, multiplying each term in the sum by the
appropriate combinatoric weight.    We thus have
the sum
\begin{eqnarray}
R^{\rm (eik)}
&\propto&
 \sum_{\{N_n / \sum N_n = j\} } \frac{j!}{N_{1}!N_{2}!\dots \overline{N_m!} \dots N_E!}
\nonumber \\
&\ & \times\ \left(\prod_{n \ne m}^{E}(p_{n}\cdot v_{m})^{N_{n}}\right)\, ,
\label{premulti}
\end{eqnarray}
where $\overline{N_m!}$ indicates that this factor is omitted in the product.
If we use the multinomial theorem, this directly simplifies to
\begin{eqnarray}
R^{\rm (eik)}
&\propto&
\left(\sum_{n\ne m}(p_{n}\cdot v_{m})\right)^{j}
\nonumber \\
&=& \quad \left(\, -p_{m}\cdot v_{m}\, \right)^{j}
\nonumber \\
&=& \quad 0\, ,
\end{eqnarray}
where in the second line we have used momentum conservation,
and in the third the assumed masslessness of $p_m$,
which implies $p_m\cdot v_m=0$.

\bigskip

\bibliographystyle{utphys}

\begin{thebibliography}{99}

\bibitem{Weinberg}
S.~Weinberg, \href{http://prola.aps.org/abstract/PR/v140/i2B/pB516_1}{{
  Phys. Rev.} {\bf 140}, B516 (1965)}.
  
  
\bibitem{Donoghue}
J.~F.~Donoghue, T.~Torma
{Phys. Rev. D} {\bf 60}, 024003 (1999).
arXiv: hep-th/9901156

\bibitem{Naculich:2011ry}
  S.~G.~Naculich and H.~J.~Schnitzer,
  JHEP {\bf 1105}, 087 (2011).
  arXiv:1101.1524 [hep-th].

\bibitem{White:2011yy}
  C.~D.~White,
  JHEP {\bf 1103} 079 (2011).
  arXiv:1103.2981 [hep-th].
  
  \bibitem{Bern:2006kd}
  Z.~Bern, L.~J.~Dixon and R.~Roiban,
  Phys.\ Lett.\  B {\bf 644}, 265 (2007)
  [arXiv:hep-th/0611086].

\bibitem{Giddings}
S.~B.~Giddings,
arXiv:1105.6359[hep-th]

\bibitem{Sterman1}
G.~Sterman, \href{http://prd.aps.org/abstract/PRD/v17/i10/p2773_1}{{
  Phys. Rev. D} {\bf 17}, 2773 (1978)}.
  
  \bibitem{Akhoury}
A.~Akhoury, \href{http://prd.aps.org/abstract/PRD/v19/i4/p1250_1}{{
  Phys. Rev. D} {\bf 19}, 1250 (1979)}; \\
  A.~Sen,
  Phys.\ Rev.\  {\bf D28}, 860 (1983).
  
  \bibitem{Libby:1978qf}
  S.~B.~Libby and G.~F.~Sterman,
  Phys.\ Rev.\  D {\bf 18}, 3252 (1978).
  
   \bibitem{Sterman:1994ce}
  G.~F.~Sterman,
  ``An Introduction to quantum field theory,''
{\it  Cambridge, UK: Univ. Pr. (1993) 572 p}.
  
\bibitem{Feynman:1963ax}
  R.~P.~Feynman,
  Acta Phys.\ Polon.\  {\bf 24}, 697-722 (1963).
  B.~S.~DeWitt,
  Phys.\ Rev.\  {\bf 160}, 1113 (1967);  
  1195;
 1239;\\
  L.~D.~Faddeev, V.~N.~Popov,
  Phys.\ Lett.\  {\bf B25}, 29-30 (1967).
 
 
\bibitem{'tHooft:1974bx}
  G.~'t Hooft, M.~J.~G.~Veltman,
  Annales Poincare Phys.\ Theor.\  {\bf A20}, 69-94 (1974);\\
  V.~E.~Korepin,
  [arXiv:0905.2175 [gr-qc]].
 
\bibitem{Capper:1973pv}
  D.~M.~Capper, G.~Leibbrandt, M.~Ramon Medrano,
  Phys.\ Rev.\  {\bf D8}, 4320-4331 (1973).
  
    \bibitem{Berends}
F.~A.~Berends and R.~Gastmans, \href{http://www.sciencedirect.com/science/article/pii/0550321375905283}{{
  Nucl. Phys. B} {\bf 88}, 99 (1975)}.
  
\bibitem{Donoghue:1994dn}
  J.~F.~Donoghue,
  Phys.\ Rev.\  {\bf D50}, 3874-3888 (1994).
  [gr-qc/9405057].
  
\bibitem{Landau:1959fi}
  L.~D.~Landau,
  Nucl.\ Phys.\  {\bf 13}, 181-192 (1959);\\
  J.\ D.\ Bjorken, doctoral dissertation, Stanford University (1959).
  
  \bibitem{Edenetal}
  R.\ J.\ Eden, P.\ V.\ Landshooff, D.\ I.\ Olive and J.\ C.\ Polkinghorne,
  ``The Analytic S-Matrix", {\it Cambridge, UK: Univ. Pr.\ (1966) 287 p}.
  
  \bibitem{Collinsbook}
J.\ Collins, ``Foundations of Perturbative QCD" {\it  Cambridge, UK: Univ. Pr.\ (2011) 636 p}.
  
\bibitem{Bauer:2000yr}
  C.~W.~Bauer, S.~Fleming, D.~Pirjol, I.~W.~Stewart,
  Phys.\ Rev.\  {\bf D63}, 114020 (2001).
  [hep-ph/0011336].

\bibitem{Goldberg}
J.~Goldberg, \href{http://prola.aps.org/abstract/PR/v111/i1/p315_1}{{
  Phys. Rev.} {\bf 111}, 315 (1958)}.
  
\bibitem{Collins:1989gx}
  J.~C.~Collins, D.~E.~Soper and G.~F.~Sterman,
  in ``Factorization of Hard Processes in QCD,'', A.H.\ Mueller, ed.,
  Adv.\ Ser.\ Direct.\ High Energy Phys.\  {\bf 5}, 1 (1988)
  [arXiv:hep-ph/0409313].
  
\bibitem{Collins:1982wa}
  J.~C.~Collins, D.~E.~Soper and G.~F.~Sterman,
  Nucl.\ Phys.\  B {\bf 223}, 381 (1983).
 
 \bibitem{Weinberg2}
S.~Weinberg, \href{http://prola.aps.org/abstract/PR/v135/i4B/pB1049_1}{{
  Phys. Rev.} {\bf 135}, B1049 (1964)}.

 \bibitem{VanN}
P.~van~Nieuwenhuizen, \href{http://prd.aps.org/abstract/PRD/v7/i8/p2300_1}{{
  Phys. Rev. D} {\bf 7}, 2300 (1973)}.

\bibitem{Giddings:2010pp}
  S.~B.~Giddings, M.~Schmidt-Sommerfeld, J.~R.~Andersen,
  Phys.\ Rev.\  {\bf D82}, 104022 (2010).
  [arXiv:1005.5408 [hep-th]];\\
  P.~Lodone, V.~S.~Rychkov,
  JHEP {\bf 0912}, 036 (2009).
  [arXiv:0909.3519 [hep-ph]];\\
  H.~J.~Schnitzer,
  [hep-th/0701217].


\end{thebibliography}

\end{document}